\documentclass[sigconf,authorversion,nonacm]{acmart}
\AtBeginDocument{%
  }

\begin{document}

\title{Yambda-5B — {A} Large-Scale Multi-modal Dataset for Ranking And Retrieval}

\author{Alexander Ploshkin}
\email{ploshkin@yandex-team.ru}
\orcid{0009-0008-4535-4204}
\affiliation{%
  \institution{Yandex}
  \country{}
}

\author{Vladislav Tytskiy}
\email{tytskiy@yandex-team.ru}
\orcid{0009-0003-3960-8689}
\affiliation{%
  \institution{Yandex}
  \country{}
}

\author{Alexey Pismenny}
\email{pismenny-alex@yandex-team.ru}
\orcid{0009-0004-0451-5260}
\affiliation{%
  \institution{Yandex}
  \country{}
}

\author{Vladimir Baikalov}
\email{deadinside@yandex-team.ru}
\orcid{0009-0009-4864-2305}
\affiliation{%
  \institution{Yandex}
  \country{}
}

\author{Evgeny Taychinov}
\email{taychinov@yandex-team.ru}
\orcid{0009-0006-2128-0701}
\affiliation{%
  \institution{Yandex}
  \country{}
}

\author{Artem Permiakov}
\email{artpermiakov@yandex-team.ru}
\orcid{0009-0009-5346-243X}
\affiliation{%
  \institution{Yandex}
  \country{}
}

\author{Daniil Burlakov}
\email{burlada@yandex.ru}
\orcid{0009-0006-7957-1712}
\affiliation{%
  \institution{}
  \country{}
}

\author{Eugene Krofto}
\email{singleton@yandex-team.ru}
\orcid{0009-0001-2627-4056}
\affiliation{%
  \institution{Yandex}
  \country{}
}

\author{Nikolay Savushkin}
\email{penguin-diver@yandex-team.ru}
\orcid{0009-0007-3791-5561}
\affiliation{%
  \institution{Yandex}
  \country{}
}

\renewcommand{\shortauthors}{A. Ploshkin et al.}

\begin{abstract}
We present \textit{Yambda-5B}, a large-scale open dataset sourced from the Yandex.Music streaming platform. Yambda-5B contains 4.79 billion user-item interactions from 1 million users across 9.39 million tracks. The dataset includes two primary types of interactions: implicit feedback (listening events) and explicit feedback (likes, dislikes, unlikes and undislikes). In addition, we provide audio embeddings for most tracks, generated by a convolutional neural network trained on audio spectrograms.

A key distinguishing feature of Yambda-5B is the inclusion of the is\_organic flag, which separates organic user actions from recommendation-driven events. This distinction is critical for developing and evaluating machine learning algorithms, as Yandex.Music relies on recommender systems to personalize track selection for users.

To support rigorous benchmarking, we introduce an evaluation protocol based on a Global Temporal Split, allowing recommendation algorithms to be assessed in conditions that closely mirror real-world use. We report benchmark results for standard baselines (ItemKNN, iALS) and advanced models (SANSA, SASRec) using a variety of evaluation metrics.

By releasing Yambda-5B to the community, we aim to provide a readily accessible, industrial-scale resource to advance research, foster innovation, and promote reproducible results in recommender systems.
\end{abstract}


\maketitle

\section{Introduction}
Modern recommender systems have become a cornerstone of the digital ecosystem, driving the success of music streaming services and short-video platforms. Their ability to personalize content directly impacts user engagement and monetization of services. However, the efficacy of these systems is critically dependent on the quality and scale of training data. Historically, recommendation algorithms have evolved from classical collaborative filtering methods to neural architectures such as DSSM \cite{huang2013learning} and DCN \citeN{wang2017deep, wang2021dcn}, followed by sequential models based on LSTM \cite{hochreiter1997long} and GRU \cite{chung2014empirical} frameworks (e.g., GRU4Rec \cite{hidasi2015session}). A breakthrough emerged with the adoption of transformers: approaches such as BERT4Rec \cite{sun2019bert4rec} and SASRec \cite{kang2018self} established new accuracy benchmarks, demonstrating the ability to capture long-term dependencies in user sessions.

Empirical studies in adjacent domains, such as computer vision (ViT \cite{dosovitskiy2020image}) and natural language processing (GPT-3 \cite{brown2020language}, Chinchilla \cite{hoffmann2022training}), confirm that scaling data and model parameters is a key driver of performance. For instance, \cite{kaplan2020scaling} formalized scaling laws, showing that NLP model performance improves polynomially with increased data and computational resources. Similar trends are observed in recommender systems \citeN{ardalani2022understanding, zhang2024scaling, zhang2024wukong}, where industrial solutions routinely leverage terabytes of data—unavailable to the academic community.

A critical barrier to research remains the limited availability of representative datasets. Commercial platforms rarely release raw data due to its strategic value, forcing researchers to rely on outdated (e.g., Netflix Prize \cite{bennett2007netflix}) and relatively small benchmarks (e.g., MovieLens \cite{harper2015movielens}, Steam \cite{kang2018self}). For example, the popular Spotify Million Playlist \cite{chen2018recsys} dataset contains only 1 million playlists, orders of magnitude smaller than real-world industrial scenarios. This creates a gap between academic experiments and practical requirements: models trained on small-scale data often lose efficacy when scaled.

To address this challenge, we introduce Yambda\footnote{\url{https://huggingface.co/datasets/yandex/yambda}} (\textbf{YA}ndex \textbf{M}usic \textbf{B}illion-interactions \textbf{DA}taset) --- one of the largest open datasets of music listening interactions, comprising 4.79 billion events (listening events, likes, dislikes, unlikes and undislikes) from 1 million anonymized users and 9.39 million tracks, collected over 11 months. Additionally, the dataset includes track metadata (duration, content embedding, artist, album) and timestamps, ensuring compatibility with modern sequential and context-aware recommendation approaches. The release of Yambda aims to democratize research in recommender systems: lowering barriers to experiment reproducibility, validating hypotheses about model scaling, and developing methods for extreme data sparsity.

The article is structured as follows: Sec.~\ref{sec:related} analyzes limitations of existing datasets, Sec.~\ref{sec:yambda} details Yambda’s collection and preprocessing methodology and presents dataset statistics, Sec.~\ref{sec:bench} defines evaluation process of the  baseline models and provide results, and Sec.~\ref{sec:future} discusses implications for research and industry.

\section{Related Work}\label{sec:related}
This section examines prominent datasets in recommender systems research that capture user-item interactions, highlighting their characteristics and limitations.

\begin{table}
  \caption{Dataset Size Comparison}
  \label{dataset_comparison}
  \begin{tabular}{lrrr}
    \toprule
    Dataset & Users & Items & Interactions \\
    \midrule
     % MovieLens-100k & 1000 & 1700 & 0.10M \\
     % MovieLens-1M & 6000 & 4000 & 1.00M \\
     MovieLens-10M & 72K & 10K & 10M \\
     MovieLens-20M & 138K & 27K & 20M \\
     MovieLens-25M & 162K & 62K & 25M \\
     MovieLens-32M & 201K & 88K & 32M \\
     Steam \cite{kang2018self} & 2.6M & 32K & 8M \\
     Netflix \cite{bennett2007netflix} & 480K & 18K & 100M \\
     Amazon Reviews 2013 \cite{mcauley2013hidden} & 6.6M  & 2.4M  & 35M \\
     Amazon Reviews 2014 \cite{mcauley2015image} & 21M & 9.9M & 83M \\
     Amazon Reviews 2018 \cite{ni2019justifying} & \underline{44M} & 15M & 233M \\
     Amazon Reviews 2023 \cite{hou2024bridging} & \textbf{55M} & \underline{48M}  & 572M \\
     Criteo 1TB Click Logs \cite{criteo2016clicklogs} & N/A & N/A & 4B \\
     % Music4All & 0.02M & 0.11M & N/A \\
     Music4All-Onion \cite{moscati2022music4all} & 119K & 109K & 253M \\
     LFM-1b \cite{schedl2016lfm} & 120K & 3.1M & 1B \\
     LFM-2b \cite{schedl2022lfm} & 120K & \textbf{51M} & 2B \\
     MLHD \cite{vigliensoni2017music} & 583K & 7M & \textbf{27B} \\
     \hline & \\[-2.2ex]
     Yambda-50M & 10K & 0.9M & 48M  \\
     Yambda-500M & 100K & 3M & 480M \\
     Yambda-5B & 1M & 9.4M & \underline{4.8B} \\
    \bottomrule
  \end{tabular}
\end{table}

\paragraph{MovieLens \cite{harper2015movielens}}

Available in six variants (100K, 1M, 10M, 20M, 25M, 32M interactions~\cite{movielens}), this dataset was first released in 1998 and remains widely adopted in academic studies due to its longevity and accessibility.
It records user-provided movie ratings on a 1--5 scale, along with optional tags.
Each interaction includes precise timestamps, and movie metadata includes titles, release years, and genre annotations.
While its simplicity facilitates matrix factorization-based approaches, the limited item pool (~10,000 movies) renders it unrepresentative of industrial-scale scenarios, where catalogs often exceed millions of items.

\paragraph{Steam}

Derived from the gaming platform, this dataset contains user reviews of games, including binary recommendations (recommend/not recommend), review helpfulness scores, and daily-level timestamps.
While text reviews provide rich contextual signals, the dataset’s focus on explicit feedback (e.g., reviews posted post-consumption) limits its utility for modeling real-time sequential interactions, a critical requirement for platforms like music or short-video streaming services.

\paragraph{Netflix}

The iconic dataset from the 2006–2009 Netflix Prize competition comprises anonymized 1–5 star ratings with associated dates.
Despite its historical significance, its sparse temporal granularity (date-only precision) and small item catalog (~17,000 movies) restrict its applicability to modern sequential recommendation tasks requiring millisecond-level precision.

\paragraph{Amazon Reviews}

Spanning four iterations since 2013, this large-scale e-commerce dataset includes product reviews, ratings, and extensive metadata (titles, descriptions, categories, prices, images).
Interactions feature millisecond-precision timestamps, crucial for modeling purchase sequences in recommendation scenarios. However, its emphasis on explicit feedback (reviews) may underrepresent implicit signals (e.g., clicks, dwell time), which dominate many real-world systems.

\paragraph{Criteo 1TB Click Logs}

A terabyte-scale dataset for click-through rate (CTR) prediction, it aggregates ad impressions and clicks.
Although its sheer size (1 billion users, 500 million banners) aligns with industrial needs, the absence of official feature documentation, timestamps, and user/item identifiers hinders reproducibility and limits its utility for sequential or context-aware modeling.

\paragraph{Music4All-Onion}

This datasets combine content-based features (lyrics, audio spectrogram analyses, video clip embeddings) with listening histories from Last.fm.
While it enables multimodal recommendation research, its limited interaction counts ($\approx$250M) and focus on content metadata rather than behavioral sequences reduce its suitability for large-scale sequential modeling.

\paragraph{LFM-1b and LFM-2b \cite{schedl2022lfm}}

Containing 1 billion and 2 billion interactions respectively, these large-scale datasets include audio features, track metadata, and user demographics.
However, licensing restrictions currently block public access, severely limiting their academic utility.

\paragraph{Music Listening Histories Dataset \cite{vigliensoni2017music}}

This dataset encompasses a comprehensive collection of user listening histories from the Last.fm platform.
While it aggregates data from a substantial user base (583K) and an extensive catalog of tracks (>7M), it currently faces accessibility limitations and remains unavailable for utilization in recommender systems research

To bridge the gap between academic research and industrial practice, we identify three critical dataset features:
\begin{itemize}
    \item Large-Scale: Datasets must approach industrial scales (millions of users/items, billions of interactions) to validate scalability and robustness of algorithms like graph neural networks or transformer-based architectures;
    \item Diverse Feedback Types: Coexistence of implicit (clicks, skips) and explicit (ratings, likes) feedback is essential for modeling complex user behavior, particularly in domains like music streaming where rapid feedback loops are predominant;
    \item Global Temporal Split: To properly evaluate recommendation algorithms datasets should provide timestamp for each event or global ordering of all events.
\end{itemize}

Current datasets often lack one or more of these attributes. For instance, MovieLens and Netflix are relatively small and contains only explicit signals, while Criteo’s lack of explicit user and item identifiers makes sequential analysis difficult. By addressing these gaps, our proposed dataset, Yambda, aims to provide a comprehensive resource for the research of next-generation recommender systems.

\section{The Yambda Dataset}\label{sec:yambda}

\subsection{Dataset Content}

\begin{table}
  \caption{Recommendation-Driven Events Ratio}
  \label{tab:is_organic_distr}
  \begin{tabular}{lrrr}
    \toprule
    Event & Total & Recommendation & Ratio \\
    \midrule
    Listen & 4,649,567,411 & 2,266,400,808 & 48.74\% \\
    Like & 89,334,605 & 37,789,576 & 42.30\% \\
    Dislike & 11,579,143 & 5,612,434 & 48.47\% \\
    Unlike & 32,944,520 & 1,651,117 & 5.01\% \\
    Undislike & 2,434,208 & 239,135 & 9.82\% \\
    \bottomrule
  \end{tabular}
\end{table}

Yandex.Music is a streaming music service that employs recommendation technologies to select relevant tracks in real time and curate personalized daily playlists. The primary user action is track listening (implicit feedback), which includes track length and listening percentage. Additionally, users can express explicit feedback by liking or disliking tracks, as well as canceling these actions (unlike and undislike). Thus, the Yambda dataset consists of five types of user-item interactions: Listen, Like, Dislike, Unlike, and Undislike.

User actions within Yandex.Music can be categorized as organic (where a user independently discovers and interacts with a track) or recommendation-driven (where interactions occur via recommendations from the service’s algorithm). The dataset includes an \texttt{is\_organic} flag for each event, enabling the distinction between organic and recommendation-driven actions. This feature is a critical aspect of Yambda, as it allows researchers to disentangle the platform’s logging policies from purely recommendation-influenced behavior. Tab.~\ref{tab:is_organic_distr} shows the ratio between organic and recommendation-driven events in the dataset.

Furthermore, Yambda provided neural embeddings derived from convolutional neural network trained in contrastive manner \cite{spijkervet2021contrastive}. These embeddings facilitate refined content-based recommendations for musical compositions and enable the application of modern approaches leveraging Semantic IDs \cite{rajput2023recommender}.

\subsection{Data Availability}

\begin{table*}
  \caption{Different Sizes of the Dataset}
  \label{tab:yambda_sizes}
  \begin{tabular}{lrrrrr}
    \toprule
    Dataset & Users & Items & Listens & Likes & Dislikes\\
    \midrule
     Yambda-50M  & 10,000    & 934,057   & 46,467,212    & 881,456    & 107,776 \\
     Yambda-500M & 100,000   & 3,004,578 & 466,512,103   & 9,033,960  & 1,128,113 \\
     Yambda-5B   & 1,000,000 & 9,390,623 & 4,649,567,411 & 89,334,605 & 11,579,143 \\
    \bottomrule
  \end{tabular}
\end{table*}

To address accessibility of the dataset, we released two subsampled variants along with the original one: Yambda-5B (full scale), Yambda-500M (1/10 subsample), and Yambda-50M (1/100 subsample), created by randomly sampling users at corresponding ratios. Tab.~\ref{tab:yambda_sizes} summarizes their key statistics.

Each variant is available in two formats:
\begin{itemize}
    \item \textit{Flat}: events are stored as tuples \texttt{(uid, item\_id, timestamp, is\_organic, ...)} in tabular form, with separate tables per interaction type;
    \item \textit{Sequential}: user histories are aggregated into timestamp-sorted lists for each interaction type, facilitating sequence-based modeling.
\end{itemize}

Moreover, for ease of use of the dataset for evaluation, we combined all event types into one \texttt{multi\_event} file.

Data are distributed using well-proven Apache Parquet format, ensuring compatibility with distributed processing frameworks (e.g., Hadoop, Spark) and modern analytical tools (e.g., Polars, Pandas). Tab.~\ref{tab:file_structure} reflects the structure and contents of the files in the dataset.

\begin{table*}
  \caption{Dataset File Structure}
  \label{tab:file_structure}
  \begin{tabular}{ll}
    \toprule
    File & Schema \\
    \midrule
    \texttt{artist\_item\_mapping.parquet} & \textcolor{violet}{artist\_id}, \textcolor{cyan}{item\_id} \\
    \texttt{album\_item\_mapping.parquet} & \textcolor{orange}{album\_id}, \textcolor{cyan}{item\_id} \\
    \texttt{embeddings.parquet} & \textcolor{cyan}{item\_id}, embed, normalized\_embed \\
    \texttt{\{size\}/listens.parquet} & \textcolor{red}{uid}, \textcolor{cyan}{item\_id}, timestamp, is\_organic, played\_ratio\_pct, track\_length\_seconds \\
    \texttt{\{size\}/likes.parquet} & \textcolor{red}{uid}, \textcolor{cyan}{item\_id}, timestamp, is\_organic \\
    \texttt{\{size\}/dislikes.parquet} & \textcolor{red}{uid}, \textcolor{cyan}{item\_id}, timestamp, is\_organic \\
    \texttt{\{size\}/unlikes.parquet} & \textcolor{red}{uid}, \textcolor{cyan}{item\_id}, timestamp, is\_organic \\
    \texttt{\{size\}/undislikes.parquet} & \textcolor{red}{uid}, \textcolor{cyan}{item\_id}, timestamp, is\_organic \\
    \texttt{\{size\}/multi\_event.parquet} & \textcolor{red}{uid}, \textcolor{cyan}{item\_id}, timestamp, is\_organic, played\_ratio\_pct, track\_length\_seconds, event\_type \\
    \bottomrule
  \end{tabular}
\end{table*}

\subsection{Acquisition And Processing}

To construct the dataset, we defined an approximately 11-month observation period and identified users who performed at least 10 target actions during the first 10 months and at least one action in the subsequent period. Then we randomly sampled 1 million users meeting these criteria and collect their interaction history for the specified period. In compliance with platform policies, all user and track data were anonymized. The dataset contains exclusively numerical identifiers for users, tracks, albums, and artists, along with their relational mappings. Event timestamps were consistently transformed as follows:
$$T'_{event} = \left[ \frac{T_{event} - T_{start}}{5} \right] \times 5$$

where $T_{start}$ denotes timestamp of the first event in the dataset. This transformation preserves temporal ordering with 5-second precision. Similarly, track length is rounded to 5-seconds precision, and listening percentage available at 1-percent granularity.

\subsection{Analysis And Statistics}

\begin{figure}[h]
  \centering
  \includegraphics[width=\linewidth]{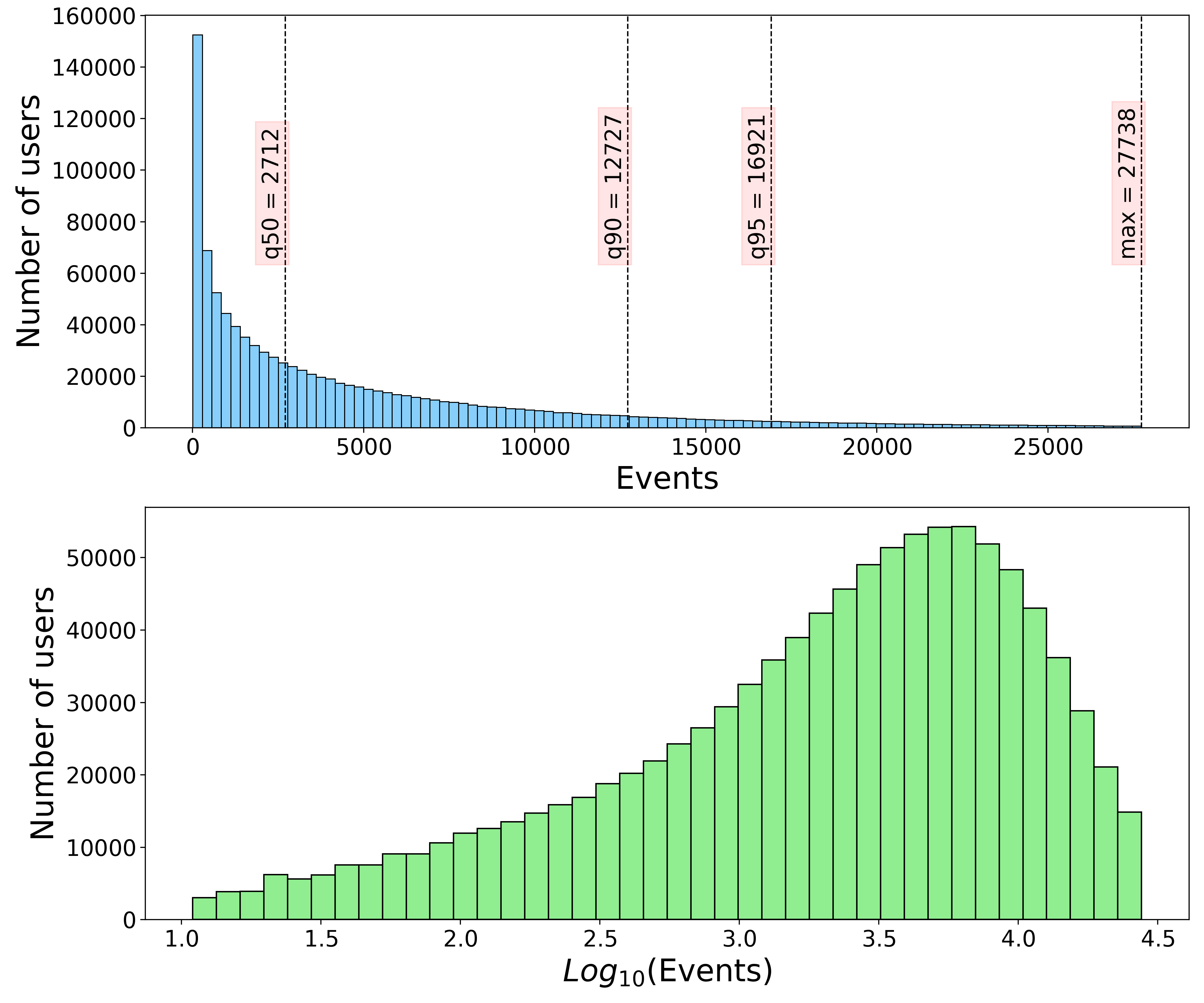}
  \caption{User History Length Distribution}
  \Description{Two graphs containing histograms of user history length. Upper is the standard histogram, lower is the log-scaled over X-axis histogram.}
  \label{fig:user_hist_len}
\end{figure}

Our dataset emphasizes user personalization through the analysis of interaction history with the service. Figure~\ref{fig:user_hist_len} illustrates the distribution of interaction history lengths across all event types for users during the dataset collection period. The median history length is comparable to the context window of modern large language models (LLMs), such as GPT-3. In addition, Tab.~\ref{tab:event_history_len} shows that most of the user history consists of implicit feedback (listens).

\begin{figure}[h]
  \centering
  \includegraphics[width=\linewidth]{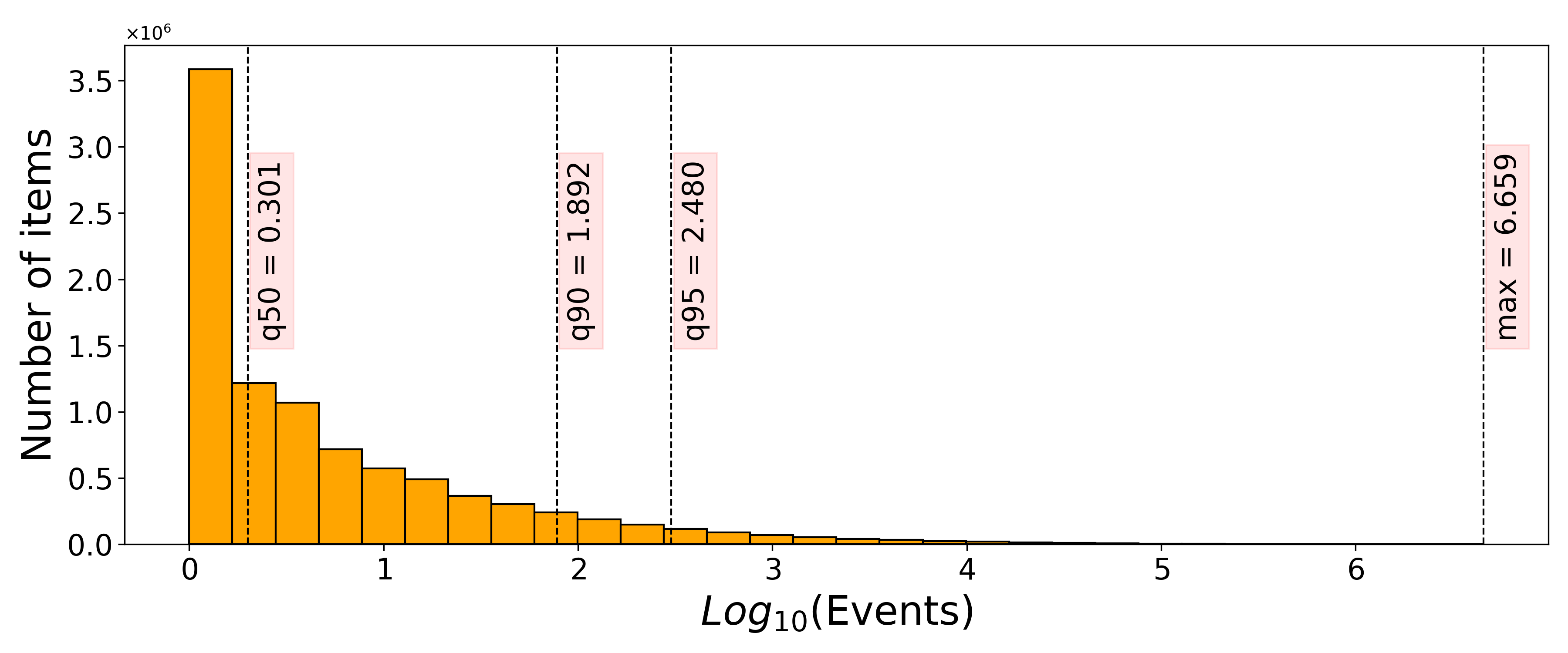}
  \caption{Item History Length Distribution}
  \Description{Log-scaled over X-axis histogram.}
  \label{fig:item_hist_len}
\end{figure}

The event distribution across items (Figure~\ref{fig:item_hist_len}) exhibits a pronounced imbalance, with a small subset of highly popular platform tracks being distinctly prominent, alongside a substantial long-tail segment of tracks exhibiting minimal (1–2) interactions.

\begin{table}
  \caption{User History Length per Event Type}
  \label{tab:event_history_len}
  \begin{tabular}{lrrr}
    \toprule
    Event & Median & p90 & p95 \\
    \midrule
    Listen & 3,076 & 12,956 & 17,030 \\
    Like & 45 & 269 & 409 \\
    Dislike & 4 & 30 & 60 \\
    Unlike & 15 & 111 & 201\\
    Undislike & 3 & 14 & 22 \\
    \bottomrule
  \end{tabular}
\end{table}

\section{Benchmarking}\label{sec:bench}

\begin{table*}
  \caption{Evaluation Results on Listen$_{+}$}
  \label{tab:eval_le}
  \begin{tabular}{llrrrrrr}
    \toprule
    Dataset & Model & NDCG@10 & NDCG@100 & Recall@10 & Recall@100 & Coverage@10 & Coverage@100 \\
    \midrule
    Yambda-50M & Random & 0.0000 & 0.0000 & 0.0000 & 0.0002 & \textbf{0.1363} & \textbf{0.7692} \\
    & MostPop & 0.0186 & 0.0249 & 0.0064 & 0.0321 & 0.0000 & 0.0002 \\
    & DecayPop & 0.0260 & 0.0323 & 0.0122 & 0.0479 & 0.0000 & 0.0002 \\
    & ItemKNN & \textbf{0.0781} & \textbf{0.0934} & \textbf{0.0373} & \textbf{0.1297} & 0.0180 & 0.0874 \\
    & iALS & 0.0407 & 0.0642 & 0.0128 & 0.0808 & 0.0044 & 0.0156 \\
    & BPR & 0.0389 & 0.0641 & 0.0139 & 0.0836 & \underline{0.0257} & \underline{0.0999} \\
    & SANSA & 0.0069 & 0.0095 & 0.0031 & 0.0137 & 0.0114 & 0.0730 \\
    & SASRec & \underline{0.0748} & \underline{0.0764} & \underline{0.0325} & \underline{0.1026} & 0.0130 & 0.0310 \\
    \hline & \\[-2.2ex]
    Yambda-500M & Random & 0.0000 & 0.0000 & 0.0000 & 0.0000 & \textbf{0.3896} & \textbf{0.9928} \\
    & MostPop & 0.0173 & 0.0237 & 0.0060 & 0.0309 & 0.0000 & 0.0001 \\
    & DecayPop & 0.0267 & 0.0340 & 0.0127 & 0.0505 & 0.0000 & 0.0001 \\
    & ItemKNN & \underline{0.0708} & \underline{0.0771} & \underline{0.0320} & \underline{0.1049} & 0.0257 & \underline{0.0917} \\
    & iALS & 0.0384 & 0.0621 & 0.0131 & 0.0811 & 0.0022 & 0.0067 \\
    & BPR & 0.0400 & 0.0652 & 0.0137 & 0.0861 & 0.0244 & 0.0758 \\
    & SANSA & --- & --- & --- & --- & --- & --- \\
    & SASRec & \textbf{0.0754} & \textbf{0.0884} & \textbf{0.0336} & \textbf{0.1240} & \underline{0.0347} & 0.0824 \\
    \hline & \\[-2.2ex]
    Yambda-5B & Random & 0.0000 & 0.0000 & 0.0000 & 0.0000 & \textbf{0.8207} & \textbf{1.0000} \\
    & MostPop & 0.0175 & 0.0239 & 0.0062 & 0.0313 & 0.0000 & 0.0000 \\
    & DecayPop & 0.0271 & 0.0348 & 0.0130 & 0.0509 & 0.0000 & 0.0000 \\
    & ItemKNN & --- & --- & --- & --- & --- & --- \\
    & iALS & 0.0388 & 0.0628 & 0.0132 & 0.0816 & 0.0010 & 0.0027 \\
    & BPR & \underline{0.0408} & \underline{0.0664} & \underline{0.0141} & \underline{0.0870} & 0.0179 & \underline{0.0508} \\
    & SANSA & --- & --- & --- & --- & --- & --- \\
    & SASRec & \textbf{0.0647} & \textbf{0.0847} & \textbf{0.0289} & \textbf{0.1214} & \underline{0.0212} & 0.0456 \\
    \bottomrule
  \end{tabular}
\end{table*}

\begin{table*}
  \caption{Evaluation Results on Like}
  \label{tab:eval_like}
  \begin{tabular}{llrrrrrr}
    \toprule
    Dataset & Model & NDCG@10 & NDCG@100 & Recall@10 & Recall@100 & Coverage@10 & Coverage@100 \\
    \midrule
    Yambda-50M & Random & 0.0000 & 0.0000 & 0.0000 & 0.0003 & \textbf{0.3677} & \textbf{0.9901} \\
    & MostPop & 0.0046 & 0.0097 & 0.0083 & 0.0222 & 0.0001 & 0.0006 \\
    & DecayPop & \textbf{0.0180} & \textbf{0.0269} & \textbf{0.0333} & \textbf{0.0651} & 0.0001 & 0.0006 \\
    & ItemKNN & \underline{0.0125} & \underline{0.0251} & \underline{0.0199} & \underline{0.0648} & \underline{0.1050} & \underline{0.3922} \\
    & iALS & 0.0078 & 0.0224 & 0.0103 & 0.0568 & 0.0110 & 0.0554 \\
    & BPR & 0.0056 & 0.0187 & 0.0073 & 0.0526 & 0.0464 & 0.1449 \\
    & SANSA & 0.0068 & 0.0203 & 0.0105 & 0.0616 & 0.0194 & 0.1601 \\
    & SASRec & 0.0100 & 0.0208 & 0.0175 & 0.0603 & 0.0211 & 0.0712 \\
    \hline & \\[-2.2ex]
    Yambda-500M & Random & 0.0000 & 0.0000 & 0.0000 & 0.0002 & \textbf{0.7272} & \textbf{1.0000} \\
    & MostPop & 0.0028 & 0.0087 & 0.0033 & 0.0215 & 0.0000 & 0.0002 \\
    & DecayPop & \textbf{0.0174} & \textbf{0.0279} & \textbf{0.0285} & \underline{0.0736} & 0.0000 & 0.0002 \\
    & ItemKNN & 0.0101 & \underline{0.0258} & 0.0151 & 0.0701 & \underline{0.1112} & \underline{0.3182} \\
    & iALS & 0.0050 & 0.0209 & 0.0064 & 0.0596 & 0.0034 & 0.0131 \\
    & BPR & 0.0071 & 0.0234 & 0.0101 & 0.0655 & 0.0628 & 0.1659 \\
    & SANSA & 0.0077 & 0.0252 & 0.0090 & 0.0681 & 0.0213 & 0.1367 \\
    & SASRec & \underline{0.0125} & 0.0237 & \underline{0.0200} & \textbf{0.0890} & 0.0392 & 0.1047 \\
    \hline & \\[-2.2ex]
    Yambda-5B & Random & 0.0000 & 0.0000 & 0.0000 & 0.0000 & \textbf{0.9838} & \textbf{1.0000} \\
    & MostPop & 0.0026 & 0.0084 & 0.0035 & 0.0238 & 0.0000 & 0.0000 \\
    & DecayPop & \textbf{0.0165} & \underline{0.0267} & \textbf{0.0281} & \underline{0.0733} & 0.0000 & 0.0000 \\
    & ItemKNN & --- & --- & --- & --- & --- & --- \\
    & iALS & 0.0052 & 0.0207 & 0.0076 & 0.0626 & 0.0012 & 0.0039 \\
    & BPR & 0.0071 & 0.0245 & 0.0106 & 0.0726 & \underline{0.0463} & \underline{0.1141} \\
    & SANSA & --- & --- & --- & --- & --- & --- \\
    & SASRec & \underline{0.0136} & \textbf{0.0348} & \underline{0.0217} & \textbf{0.1026} & 0.0224 & 0.0547 \\
    \bottomrule
  \end{tabular}
\end{table*}

To establish a foundation for utilizing this dataset as a benchmark for novel approaches in recommender systems, we implemented several baseline algorithms and proposed an offline evaluation procedure closely aligned with industrial recommendation system practices. The code is available in our GitHub repository.

\subsection{Evaluation Process}
Typically, academic studies report offline evaluations due to the lack of access to online experiments. Currently, two offline evaluation schemes are most prevalent:
\begin{itemize}
  \item Leave-One-Out (LOO), where the last positive interaction of each user is excluded from the training set and used for prediction. A key limitation of this approach is the violation of temporal dependencies, as the model may inadvertently utilize more recent interactions during training than those present in the test set;
  \item Global Temporal Split (GTS), where the entire dataset is partitioned into training and test sets based on timestamps. While this preserves temporal consistency, it risks including users absent from the training data, whose interaction histories are unknown at the start of the test period.
\end{itemize}

To better approximate real-world production environments, we adopted a modified GTS scheme:
\begin{itemize}
  \item \textit{Train}: ~300 days
  \item \textit{Gap}: 30 minutes
  \item \textit{Test}: 1 day
\end{itemize}

A 30-minute gap between training and test sets was introduced to exclude interactions used neither for training nor evaluation. This mimics the latency between model training and deployment in industrial systems. Additionally, to ensure consistency, users with empty interaction histories at the start of the test period were discarded.

All model parameters and user states were frozen at the beginning of the test period. A trade-off exists here: true real-time recommendation systems update user states with delays of tens of seconds, but replicating this would significantly complicate evaluation and increase computational costs. By limiting the test period to one day and freezing user states, we approximate systems with daily offline updates for model and user state refreshes—a reasonable compromise for practical benchmarking.

\subsection{Baselines}

We selected well-established algorithms as baselines:
\begin{itemize}
  \item \textit{MostPop}: Recommendations based on overall item popularity.
  \item \textit{DecayPop} \cite{ji2020re}: A time-decayed variant of \textit{MostPop}, better utilizing emergent popularity.
  \item \textit{ItemKNN} \cite{sarwar2001item} is a neighborhood-based collaborative filtering approach that generates recommendations by identifying items similar to those a user has previously interacted with. It employs similarity metrics (e.g., cosine or Pearson correlation) to compute item-item affinity, leveraging the assumption that users prefer items analogous to their historical preferences.
  \item \textit{iALS} \cite{hu2008collaborative} is a matrix factorization method optimized for implicit feedback datasets. It decomposes the user-item interaction matrix into latent user and item factors, regularized to prevent overfitting. The alternating least squares solver ensures scalability and efficiency, making it suitable for large-scale recommendation tasks.
  \item \textit{BPR} \cite{rendle2012bpr} is a pairwise ranking optimization framework designed for implicit feedback. It learns personalized rankings by maximizing the posterior probability that a user prefers observed items over unobserved ones. BPR employs a triplet loss to model user preferences through stochastic gradient descent, emphasizing personalized ordinal relationships.
  \item \textit{SANSA} \cite{spivsak2023scalable} is a scalable variant of EASE \cite{steck2019embarrassingly} framework, designed to address computational bottlenecks in large-item datasets. By relaxing the symmetry constraint of EASE’s item-item similarity matrix and introducing approximate training strategies, SANSA reduces time and memory complexity while retaining high recommendation accuracy, making it practical for industrial-scale applications.
  \item \textit{SASRec} \cite{kang2018self} is a transformer-based sequential model that leverages self-attention to model user behavior sequences. It dynamically assigns attention weights to past interactions, capturing temporal dependencies and highlighting influential items. SASRec excels in scenarios requiring precise modeling of evolving user preferences over time.
\end{itemize}

\subsection{Results}
For the analysis of the algorithms, we employed well-established metrics, including NDCG@$k$ ($k\in\{10,100\}$) to assess ranking quality, Recall@$k$ to evaluate candidate generation performance, and Coverage@$k$ to quantify the recommendation system’s comprehensiveness in representing the item collection. For our evaluation Coverage is calculated using the following formula:
$$\text{Coverage@}k = \frac{|\bigcup_{u \in U}{R(u, k))}|}{|I|},$$
where $U$ and $I$ denote sets of users and items respectively, $R(\cdot, \cdot)$ is a ranking function which returns top relevant items for the specified user.

We evaluated algorithm performance under two feedback setups: positive listening events, or Listen$_+$ (implicit feedback) and Like (explicit feedback). To produce Listen$_+$ we used 50\% of the track duration as the listening threshold.
Additionally, we tested the algorithms on smaller subsets of the dataset. Hyperparameters for DecayPop, ItemKNN, iALS, BPR, and SANSA were tuned using the GTS scheme. For SASRec, hyperparameter optimization was deferred due to computational constraints. To find optimal hyperparameters, we reserved one day from the training set for validation, maintaining the 30-minute gap for consistency. The NDCG metric, widely used for ranking evaluation, guided the optimization process via the OPTUNA~\cite{akiba2019optuna} framework.

Final results, obtained by training models on the full training set with tuned hyperparameters, are summarized in Tab.~\ref{tab:eval_le} and Tab.~\ref{tab:eval_like}. Metrics for ItemKNN and SANSA are unavailable at larger dataset scales due to their computational intractability within practical time constraints. We observed that the top-2 algorithms by ranking metrics in the Listen$_+$ scenario consistently included ItemKNN and SASRec across all dataset scales. Conversely, the DecayPop algorithm demonstrated superior performance in the Like scenario, establishing itself as the most effective ranking method despite its simplicity.

\section{Conclusions And Future Work}\label{sec:future}

A large-scale dataset containing 4.79 billion user-item interactions has been publicly released to advance research in recommender systems. This resource is distinguished by three critical features: (1) high-fidelity audio embeddings derived from spectral analysis of tracks, enabling content-aware recommendation approaches; (2) a binary is\_organic flag annotating interactions to differentiate organic user behavior from algorithmically influenced actions; and (3) timestamp granularity aligned with the Global Temporal Split evaluation protocol, which was rigorously implemented to prevent temporal data leakage during baseline model benchmarking.

Methodologically, the dataset’s design addresses longstanding limitations in academic research by mirroring industrial-scale conditions. The evaluation framework employed in this work demonstrates that conventional collaborative filtering methods (e.g., Matrix Factorization) exhibit degraded performance when applied to scenarios requiring real-time interaction processing, highlighting the necessity of sequence-aware architectures like Transformers.
Future Directions

Three primary research avenues are proposed to extend this work:

\paragraph{Enhanced Temporal Evaluation Protocol}
The Global Temporal Split scheme will be refined to better emulate industrial recommender systems through:
\begin{itemize}
    \item Incremental user state updates via streaming data pipelines;
    \item Periodic retraining of lightweight algorithms (e.g., iALS, BPR) at fixed intervals;
    \item Simulation of cold-start scenarios through dynamic user/item inclusion.
\end{itemize}

\paragraph{Multi-Modal Recommendation Paradigms}
The dataset’s multi-modal nature—combining behavioral sequences, audio embeddings, and track metadata—will be leveraged to investigate:
\begin{itemize}
    \item Cross-modal fusion techniques for hybrid recommender architectures;
    \item Graph neural networks exploiting artist-album-track relationships;
    \item Contrastive learning frameworks aligning audio features with user preferences.
\end{itemize}

\paragraph{Organic vs. Algorithmic Interaction Analysis}
A comprehensive analysis of behavioral differences between organic and recommendation-driven interactions will be conducted, focusing on:
\begin{itemize}
    \item Temporal patterns in user engagement decay post-recommendation;
    \item Bias propagation in feedback loops induced by algorithmic curation;
    \item Metrics quantifying the diversity and serendipity of organic discovery.
\end{itemize}

This dataset is positioned to serve as a foundational resource for bridging the gap between academic experimentation and industrial deployment. By democratizing access to web-scale interaction data while preserving privacy through rigorous anonymization, the work aims to catalyze innovation in sequential recommendation, fairness-aware algorithms, and multi-modal personalization. All artifacts, including preprocessing code and baseline implementations, have been standardized to ensure reproducibility across experimental setups.

\bibliographystyle{ACM-Reference-Format}
\bibliography{references}

\end{document}